\documentstyle[prd,aps,eqsecnum,epsfig]{revtex}
\tighten
\begin{document}
\draft
\twocolumn[\hsize\textwidth\columnwidth\hsize\csname
@twocolumnfalse\endcsname

\title{Notes on Black Hole Phase Transitions}
\author{G. J. Stephens $^*$ $^1$ $^2$ and 
B. L. Hu $^{**}$ $^2$}

\address{$^1$ Theoretical Astrophysics T-6 MS-B288, Los Alamos National Laboratory \\
Los Alamos, New Mexico 87545}

\address{$^2$ Department of Physics, University of Maryland\\
College Park, Maryland 20742-4111}

\date{\today}
\maketitle

\begin{abstract}
In these notes we present a summary of existing ideas about phase transitions of black hole
spacetimes in semiclassical gravity and  offer some thoughts on three possible scenarios or mechanisms by 
which  these
transitions could take place.  
We begin with a review of the thermodynamics of a black hole system
and emphasize that the phase transition is driven by the large entropy of the black hole
horizon. Our first theme is illustrated by a quantum {\it atomic} black hole system, generalizing to finite-temperature a model 
originally offered by Bekenstein.  In this equilibrium atomic model, the black hole phase transition is realized as the abrupt excitation
of a high energy state, suggesting analogies with the study of two-level atoms.  
Our second theme argues that the black hole system shares similarities with the defect-mediated 
Kosterlitz-Thouless transition in condensed matter.  These similarities suggest 
that the black hole phase transition may be more fully understood by focusing upon the dynamics
of black holes and white holes, the spacetime analogy of vortex and anti-vortex topological defects.  Finally, we compare
the black hole phase transition to another transition driven by an (exponentially) increasing density of 
states, the Hagedorn transition first found in hadron physics in the context of dual models or the old string theory. 
In modern string theory the Hagedorn transition is linked by the Maldacena conjecture 
to the Hawking-Page black hole phase transition in Anti-de Sitter (AdS ) space, as observed by
Witten.  Thus, the dynamics of the Hagedorn transition may yield insight into the dynamics of the black 
hole phase transition.  We argue that characteristics of the Hagedorn transition 
are already contained within the dynamics of {\it classical} string systems. Our third theme 
points to carrying out a full nonperturbative and nonequilibrium analysis of the large N behavior 
of classical $SU(N)$ gauge theories to understand its Hagadorn transition. By
invoking the Maldacena conjecture we can then gain valuable insight into black hole phase transitions in AdS space. 
\end{abstract}

\pacs{$^*$ electronic address: gstephen@physics.umd.edu \hfill $^{**}$ electronic address: hub@physics.umd.edu\\
PACS numbers: 04.70.Dy,04..60-m,05.70.Fh \hfill UMDPP-01-034 LA-UR-01-0749}

\vskip2pc]

\bigskip
\section{Introduction}
Thermal fluctuations can induce  
phase transitions in which the zero-temperature degrees of freedom are reorganized into a 
qualitatively different form.  While there are many familiar examples such as the boiling of water,
phase transitions also occur in more exotic systems like spacetime geometry.
In these notes (based partly on Chapter 5 of the dissertation work of GJS \cite{stephensdiss}),
we study a spacetime phase transition evident through the spontaneous formation of
a black hole in equilibrium with a thermal environment.  We are interested in two important questions:
\begin{quote}
1. Why does a black hole phase transition occur?

2. By what dynamical scenario does the phase transition take place?
\end{quote}
We gather here some of our recent thoughts and propose some ideas which have 
the potential of providing partial answers to these questions.

The abstract nature of the black hole phase transition is a departure from the more physical
systems familiar to us from ordinary experience.  
No laboratory is yet equipped with the tools necessary to
probe the formation of a black hole in a phase transition of thermal spacetime, although such situations
may have existed in the very early universe.  Here, 
we want to see how far our understanding of more accessible physical systems can be applied to
similar processes in spacetimes. A more ambitious intention is to use our knowledge gained from the study of quantum and classical 
phase transitions to peer into the complicated workings of general relativity and semiclassical gravity.
General relativity is a highly nonlinear theory and comparatively little is known about its detailed behavior away
from exact solutions.  For example, in the extremely energetic environment of the early universe,  
general relativity may exhibit a disordered phase dominated by black holes, wormholes, geons and other 
nonperturbative gravitational excitations.  
At even higher energies near the
Planck scale it has
long been speculated that spacetime appears as a foam or froth \cite{wheeler:1957}.  Indeed, the study of
Planck scale fluctuations including gravitational bubbles \cite{hawking:1978,hawking:1980,warner:1982},
baby universes \cite{hawking:1988,coleman:1988,coleman:1991}, 
virtual black holes \cite{hawking:1996} and black hole pair creation \cite{garfunkle:1994,dowker:1994,hawking:1995,busso:1995,busso:1996}
has been given a solid grounding by Hawking and his associates. 
Unfortunately, while the picture of
spacetime foam is vivid, it is very hard to realize in quantitative terms: in part because a theory
of quantum gravity does not yet exist. However, it is our hope that understanding the phase structure of 
semiclassical gravity provides at least an angle on the attributes of the transition to
quantum gravity.  

Black holes bear some similarity to topological defects \cite{volovik:1995}.  They are both stable, nonperturbative 
solutions of the classical theory.  Both black holes and topological defects carry with them a 
remnant of the high temperature phase; symmetry is restored in the core of topological defects and 
we are likely to find a quantum phase of spacetime in the high-curvature region near the black hole singularity. 
In addition, as we will show in these notes,  the black hole phase transition is qualitatively
similar to the defect-mediated Kosterlitz-Thouless phase transition in condensed matter and the Hagedorn transition
in string systems.

The analogy between black holes and topological defects is only a partial reflection of a deeper
dialectical relationship existing between condensed matter physics, including statistical 
mechanics and hydrodynamics and the so-called ``fundamental'' physics such as 
quantum gravity and particle physics, which refers to the study of the `basic' constituents of spacetime
and matter. The former is understood in broader terms as a study of  the 
complex organization in structure and interactions of their constituent 
elements, whether they be atoms and molecules in condensed matter, or strings in the formation of spacetime.  
In this light cosmology is closer to condensed matter physics than 
elementary particle physics  \cite{hu:1995a,hu:1995b,hu:1996a,hu:1999a,volovik:2000,zurek:1996,smolin:1995}.   Condensed matter systems  of atoms and molecules are usually easier to understand than
their high energy counterparts such as strings or geons,  and we can exploit these analogies to illuminate the behavior of systems which are 
otherwise experimentally and theoretically intractable. For example, sonic black holes in superfluids such as Bose-Einstein 
condensates may provide a testable model of black hole Hawking radiation \cite{unruh:1981,unruh:1995,jacobson:1991,garay:2000}. 
The experimental tests of physics at high 
energy scales are (and are likely to remain) relatively few.  In this environment, 
condensed matter analogs are useful in aiding the probe of fundamental physics, of relevance here being
semiclassical gravity.

In Sec.~\ref{sec:thermo} we review the thermodynamics
of the black hole phase transition, focusing on the calculation of the semiclassical free energy of a black hole
in thermal equilibrium, represented by the work of Whiting and York \cite{york:1986,whiting:1988}.  
In Sec.~\ref{sec:atom} we review an atomic model of the black hole, first
introduced as a quantum model of black hole microstates by Bekenstein \cite{bekenstein:1997}.  
In original work we
use this model to 
provide a statistical mechanics of the phase transition and highlight the important role of black hole
entropy.  In Sec.~\ref{sec:dynamics} we discuss the nonequilibrium dynamics of the black hole phase transition. 
We point out the inadequacy of the homogeneous nucleation theory of first-order phase transitions and explore  
examples from condensed matter (Kosterlitz-Thouless transition) and string theory (Hagadorn transition) which appear qualitatively
similar to the black hole system.  
A summary is provided in Sec.~\ref{sec:sum}.  Unless otherwise noted, 
we use Planck units for which $\hbar=G=c=1$.

\section{Spacetime thermodynamics}
\label{sec:thermo}
For normal matter, gravitational interactions are universally attractive and
a self-gravitating system is fundamentally unstable to collapse.   
For example, a nonrelativistic ideal homogeneous fluid with density $\rho$ and 
sound speed $v_s$ is unstable to long wavelength density perturbations.
Perturbations with wave vector 
\begin{equation}
k_J<\sqrt{\frac{4\pi \rho}{v_s^2}}
\end{equation}
will grow exponentially.  This is the Jeans instability.  Since gravity cannot be screened, gravitational
instabilities remain for a system in thermal equilibrium.  Compress an ideal, isothermal, self-gravitating gas 
below a critical volume and the gas will collapse. 

Quantum effects induce additional instabilities to a classical thermal gravitational 
system. In particular, hot, flat space is unstable to the (quantum) nucleation of black holes \cite{gross:1982}. 
Using the techniques of Euclidean quantum gravity, the nucleation of black holes was identified through the discovery of
a Schwarzschild instanton contributing an imaginary piece to the free energy of the system.  The 
nucleation rate is maximum for a black hole with mass $M=\frac{1}{8\pi T}$ and is (approximately) given by 
\begin{equation}
\Gamma \sim T^5 \exp {\left (-\frac{1}{16\pi T^2}\right )}.
\end{equation}
A black hole nucleated at temperature $T$ is in unstable equilibrium with a thermal environment of the same temperature.  If, in a fluctuation,  
the black hole absorbs a small amount of radiation, its Hawking temperature decreases (black holes generally
have negative specific heat).  As the black hole grows it becomes colder still, 
absorbing more radiation and eventually engulfing the system.  
Thus, although $\Gamma$  is small except near the Planck scale, the negative specific heat of nucleated black holes 
renders the canonical ensemble of hot, flat space ill-defined.

Black hole systems can be rendered thermodynamically stable with the addition of
special boundary conditions or when placed in spacetimes with
a negative cosmological constant \cite{hawking:1983}.  In the following we fix the temperature $T$ on an 
isothermal boundary of radius $r$ containing
a black hole of mass $M$.  In equilibrium,
the Hawking temperature measured on the boundary must equal the
boundary temperature,
\begin{equation}
\label{eq-bhT}
T(r)=\frac{1}{8\pi M} \frac{1}{\sqrt{1-\frac{2M}{r}}}.
\end{equation}
Eq.~(\ref{eq-bhT}) admits two real, nonzero solutions for the mass: a smaller, 
unstable black hole with mass $M_1$ and a larger, 
stable black hole with mass $M_2$,
\begin{eqnarray}
\label{eq-bhm}
M_1 &\simeq& \frac{1}{8\pi T} \left[1+\frac{1}{8\pi rT} \right],  \\  
M_2 &\simeq& \frac{r}{2}\left[1-\frac{1}{(4\pi rT)^2} \right ].
\end{eqnarray}
The isothermal boundary renders $M_2$ thermodynamically stable because of the temperature redshift.  A fluctuation
that increases $M_2$ also {\it increases} the temperature of the black hole as measured on the boundary,
giving $M_2$ a positive specific heat.  Surprisingly, for 
\begin{equation}
\label{eq-bhTc}
T<T_c=\frac{\sqrt{27}}{8\pi r}
\end{equation}
no real value $M$ can solve Eq.~(\ref{eq-bhT}) and no black hole can exist in the box.  It therefore
appears that as the temperature on the boundary is increased from
$T=0$, a phase transition to a black hole spacetime occurs at $T=T_c$.

To further elucidate the thermodynamics of the black hole system and the nature of any potential phase transition
we consider the canonical partition function
defined through an Euclidean path integral \cite{hawking:1979},
\begin{equation}
Z[\beta]=\int D[g]e^{-\frac{I_E[g]}{\hbar}},
\end{equation}
where we have temporarily restored the $\hbar$ dependence in anticipation of the semiclassical limit.
The Euclidean action is obtained from the Lorentzian Einstein-Hilbert action (with boundary
terms) through the Wick rotation, $t\rightarrow -i\tau$.  The functional integration is taken over 
real Euclidean metrics, periodic in imaginary time coordinate $\tau$ with period equal to the inverse
temperature $\beta$.  Apart from artificial toy models, the full functional integral is intractable.
However, in the semiclassical limit $(\hbar \rightarrow 0)$, the integrand is highly peaked around
metrics that minimize the classical Euclidean action.  In the semiclassical limit we evaluate the
partition function for the system consisting of a single black hole in a finite cavity with boundary
topology $S^1 \times S^2$ (the partition function beyond the semiclassical approximation
was considered in \cite{whiting:1988}).  The action is
\begin{equation}
\label{eq-ebha}
I_E=\frac{1}{16\pi} \int R\sqrt{g}d^4x + \frac{1}{8\pi} \oint K\sqrt{\gamma}d^3x,
\end{equation}  
where $K$ is the trace of the extrinsic curvature of the boundary and $\gamma$ is
the determinant of the induced three-metric.  The free energy of a single Schwarzschild black hole of mass $M$
within the cavity is given by $F=\beta^{-1}I_E$ where the Euclidean action Eq.~(\ref{eq-ebha}) is evaluated for the metric
\begin{equation}
ds^2=\left(1-\frac{2M}{r} \right)d\tau^2 + \left ( 1-\frac{2M}{r} \right )^{-1} dr^2 + r^2d\Omega^2.
\end{equation}
and  $\tau$ is a Euclidean time coordinate with period $\beta$.
Normalized so that $F=0$ when $M=0$, the semiclassical free energy is 
\begin{equation}
\label{eq-bhf}
F(M,r,T)=r-r\sqrt{1-\frac{2M}{r}}-4\pi M^2 T.
\end{equation}
\begin{figure}
\begin{center} 
\psfig{figure=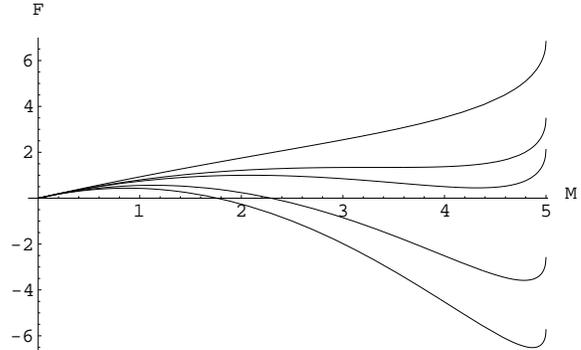,width=3.0in,angle=0}
\caption{Plot of the free energy F vs. mass M of the black hole system.} 
\label{fig:bhF}
\end{center}
\end{figure}
In Fig.~\ref{fig:bhF} we plot the free energy $F(M)$ at various temperatures for a system with box size $r=10$.  
>From the top down, the first curve is for 
temperature $T=0.01$, below the critical temperature $T_c=0.021$.  The next
curve is for $T=T_c$.  The lower three curves are for $T=0.25$, $T=0.4$ and $T=0.5$ respectively.
For temperatures below $T_c$ no black hole is present.  At $T=T_c$ an extremum
appears at $M=M_1=M_2$ (where the free energy is flat). For temperatures above $T_c$ there are two extrema, 
the unstable black hole with mass $M_1$ and the stable black hole with mass $M_2$, in agreement with the simple 
arguments at the beginning of the section.  Figure \ref{fig:bhF} also suggests that the transition to
a black hole spacetime above $T_c$ occurs discontinuously.  Above $T_c$ a finite mass black hole can be nucleated
in equilibrium with the walls of the box.

The nucleation of a stable black hole at $T=T_c$ does {\it not} necessarily signal a phase transition.   
A system in thermodynamic equilibrium always resides in a state of lowest free energy. A phase transition may occur only if the free energy of the system with the black hole is lower than the free energy without the black hole.  
At temperatures below $T_c$ no black hole is present
and the free energy is approximately that of a box filled with thermal gravitons (hot, flat space),
\begin{equation}
F_{hfs}\sim -T^4 r^3.
\end{equation}
The free energy of hot flat space is  negative.  Therefore a phase transition from hot flat spacetime to a black hole
spacetime can only occur at temperatures $T>T_c$ for which $F(M_2) < F_{hfs}$.  In Fig.~\ref{fig:bhF} 
the temperature is low enough that $F_{hfs} \approx 0$ and a black hole phase transition occurs when
$F(M_2)<0$.  

The nucleation of a black hole in equilibrium with the walls of the box is very different from 
the result of classical gravitational collapse. If the box is uniformly filled with 
massless thermal radiation, we approximate the collapse temperature
as the temperature for which the Schwarzschild radius for the thermal energy of the radiation 
is equal to the box size,
\begin{equation}
\label{eq-ctemp}
T_{collapse}^4  r^3\sim r.
\end{equation}
Thus $T_{collapse} \sim \frac{1}{\sqrt{r}}$, qualitatively distinct from the nucleation 
temperature.  

The phase transition also does not occur at a high temperature characteristic of quantum gravity.
In fact, as the size of the box is increased the critical temperature decreases and is arbitrarily low
for arbitrarily large boxes.  The stable black hole formed at temperatures above $T_c$ is large, 
with a mass $M_2$ on the order of the size of the box.  These 
considerations seem to be paradoxical.  How can such a large energy ($E\sim M$) fluctuation actually 
{\it lower} the free energy of the system at such low temperatures $T\ll M$?  
The answer lies in the enormous entropy black holes hold within their horizon.  The free energy
results from a competition between the internal energy and 
the entropy, $F=E-TS$. For a black hole, $E=M$ and the entropy is proportional to the 
area of the horizon, $S=4\pi M^2$.
Because the entropy is growing with mass faster than the energy, there is always a 
critical temperature above which the entropy completely
compensates for the energy cost of making a black hole and the black hole spacetime is the lowest 
free energy state.  
\section{Equilibrium black hole atoms}
\label{sec:atom}
The black hole phase transition is {\it entropically} driven.  Therefore,  an understanding of the nature of black hole entropy may offer potential insight into the details of the transition. In this 
section we study a toy model for black hole microstates, focusing on their implications for the thermal
black hole system.   

The origin of black hole entropy has been an outstanding problem since 
Bekenstein first introduced the concept \cite{bekenstein:1973}.  A complete resolution likely requires
a consistent theory of quantum gravity, which has so far proved elusive.  However,
just as semiclassical reasoning such as the Bohr model was important in the early 
development of quantum theory and the interpretation of atomic spectra, a similar approach 
may be fruitful in understanding some of the microscopic features of black hole entropy \cite{bekenstein:1997}.    
It is not our intention to survey the large number of semiclassical black hole models. However common to many 
is the quantization of the horizon area into equally spaced levels (see \cite{kastrup:1997} and references therein),
\begin{equation}
A=\alpha n; \;\;\;\;\;\; n=1,2\ldots 
\end{equation}
where $\alpha$ is dimensionless but as yet unspecified and the area is given in units
of Planck area.  For a black hole of mass $M$ the entropy,
\begin{equation}
\label{eq-entropy}
S=\frac{A}{4}=4\pi M^2,
\end{equation}
and thus the mass of the black hole is also quantized, 
\begin{equation}
\label{eq-masslevel}
M =\gamma \sqrt{n},
\end{equation}
where $\gamma= \sqrt{\frac{\alpha}{16\pi}}$.  If the energy levels have degeneracy
$g(n)$, the black hole entropy may simply count the number of available
microstates for a fixed energy level $n$, $S=\ln{g(n)}$.  In this case,
\begin{equation}
\label{eq-degeneracy}
g(n)=e^{\frac{\alpha n}{4}}.
\end{equation}  
Since $g(n)$ must also be an integer, $\alpha$ is restricted,
\begin{equation}
\label{eq-alpha}
\alpha=4\ln{k}; \;\;\;\;\;\;\; k=2,4\ldots
\end{equation}
Equations (\ref{eq-masslevel}), (\ref{eq-degeneracy}) and (\ref{eq-alpha}) define a semiclassical black hole 
model in analogy with
the Bohr model of an atomic system. If $k=2$ the large degeneracy $g(n)$ can be thought of as the number of ways to
make a black hole in the nth level by starting in the ground state \cite{mukhanov:1986}, though there are
other interpretations \cite{danielsson:1993,kastrup:1999}. Our interest in the atomic black hole model is its behavior in 
thermal equilibrium.    

\begin{figure}
\begin{center}
\strut\psfig{figure=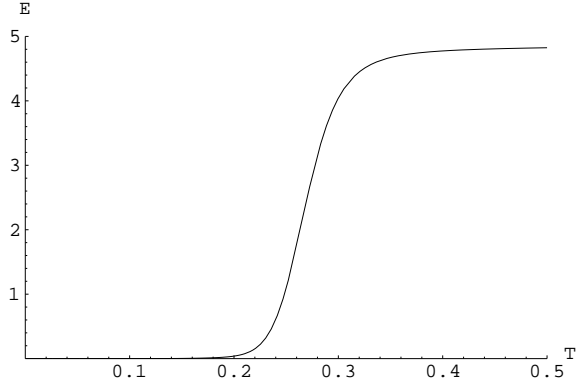,width=3.0in,angle=0}
\caption{Plot of the average energy $E$ vs. temperature $T$ for the black hole atom.} 
\label{fig:abhfig1}
\end{center}
\end{figure}

\begin{figure}
\begin{center}
\strut\psfig{figure=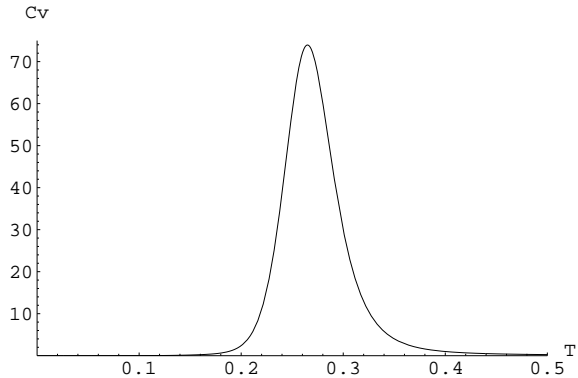,width=3.0in,angle=0}
\caption{Plot of the specific heat $C_v$  vs. temperature $T$ for the black hole atom.} 
\label{fig:abhfig2}
\end{center}
\end{figure}

\begin{figure}
\begin{center}
\strut\psfig{figure=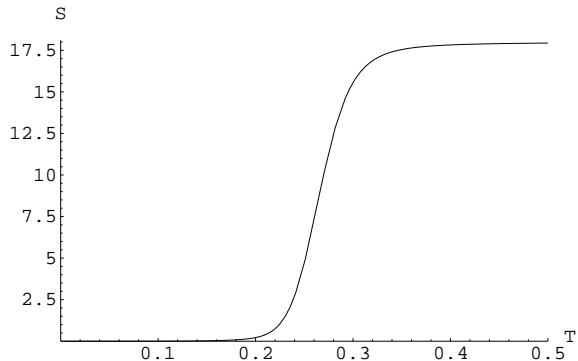,width=3.0in,angle=0}
\caption{Plot of the entropy $S$ vs. temperature $T$ for the black hole atom.} 
\label{fig:abhfig3}
\end{center}
\end{figure}

The partition function for the quantum black hole (QBH) atom in the canonical ensemble is
\begin{equation}
\label{eq-abhZ}
Z[T]=\sum_{n=0}^{\infty} k^n e^{-\frac{ \gamma \sqrt{n}}{T}}.
\end{equation}
Without  modification, the sum in Eq.~(\ref{eq-abhZ}) does not converge.  Convergence is obtained upon 
analytic continuation but the partition function acquires an imaginary piece \cite{kastrup:1999}.  In light of 
the discussion
of the previous section this is not at all surprising.  Without either invoking special boundary conditions or special
spacetimes, the thermodynamics of black holes is not well-defined.  In fact, Im(Z) is due precisely to
the nucleation of black holes.  To obtain a well-defined and real partition function we take a new approach
and place the quantum black hole atom into a box of radius $r$.  The box is realized as a sharp cutoff 
in the energy levels accessible to the black hole,
\begin{equation}
n_{max}=\frac{r^2}{4\gamma^2}.
\end{equation}
This is reasonable as the box acts to remove energy levels with Schwarzschild radius larger than $r$.  
The canonical partition function for the
quantum black hole
atom in a box is
\begin{equation}
\label{eq-boxabhZ}
Z[T]=\sum_{n=0}^{\frac{r^2}{4\gamma^2}} k^n e^{-\frac{\gamma \sqrt{n}}{T}}.
\end{equation}
In Figs.~\ref{fig:abhfig1}, \ref{fig:abhfig2} and \ref{fig:abhfig3} we plot the average energy, specific heat and 
entropy as a function of temperature for a quantum black hole atom in a box with radius $r=10$.
The plots were made with rescaled variables $T\rightarrow \gamma T$, $r\rightarrow \gamma r$
and for the particular choice $k=2$. All thermodynamic quantities were calculated 
using the partition function Eq.~(\ref{eq-boxabhZ}).
Inspection of Figs.~\ref{fig:abhfig1}-\ref{fig:abhfig3} reveals a sharp transition from the ground state to
a highly excited state in the black hole atomic system.   This is reminiscent of the phase transition behavior that we saw in
the previous section.  To quantify this behavior consider the effective Boltzmann factor of level $n$,
\begin{equation}
f(n)=-\frac{\gamma \sqrt{n}}{T}+n\ln{k}.
\end{equation}
For small $n$, $f(n)$ is negative and higher energy levels are suppressed, as is usually the case.
However, since the degeneracy grows as $n$, there is always an energy level $n_c(T)$ defined by $f(n_c)=0$
beyond which degeneracy compensates for the higher energy.  Levels beyond $n_c$ are {\it enhanced}, not suppressed.  
At low $T$, $n_c(T) > n_{max}$ and these enhanced levels are not part of the allowed spectrum.  
As T increases $n_c(T)$ decreases and a transition to the higher levels
occurs when $n_c(T) = n_{max}$.  If we adopt this picture of the black hole phase transition the critical temperature is
\begin{equation}
\label{eq-bhaTc}
T_c=\frac{1}{2\pi r}.
\end{equation}
The average energy at the critical point is
\begin{equation}
\label{eq-bhaM}
E(T_c) \sim e^{f(n_c(T_c))}\gamma \sqrt{n_{max}}=\frac{r}{2}.
\end{equation}
These equations reveal two important details.  First, although the numerical factors are slightly
different they have the same qualitative structure as their thermodynamic counterparts,
Eqs.~(\ref{eq-bhTc}) and (\ref{eq-bhm}).  Second, neither equation contains the one free parameter $k$  
of the black hole atom.  In fact Eqs.~(\ref{eq-bhaTc}) and (\ref{eq-bhaM}) are largely independent of the 
details of the spectrum of the black hole atom and apply even when the horizon area is quantized into nonuniform levels.

At first glance, our study of the quantum black hole atom in thermal equilibrium appears to only slightly 
advance our understanding of the black hole phase transition.  We have simply reaffirmed the fact that a
system with finite energy (limited by the boundary conditions) and obeying a thermodynamic relation $S=4\pi E^2$ will
always have a transition to the highest allowed energy state, the details of which are largely model independent.  
But in principle the quantum black hole atom allows us to go farther. We can now imagine treating dynamical
processes such as emission and absorption much as we do with more common atomic systems.  In fact only
two levels, the ground state and the highest allowed excited state, are significantly populated below
and above the transition and we may further approximate our system as a quantum two-level atom (2LA), a 
popular system of study. Since the quantum black hole transition occurs as the excitation of a high energy state,
the dynamics of this atomic excitation approximate
the dynamics of the black hole phase transition.  

Our analysis of the quantum black hole atom in thermal equilibrium underscores the entropic nature of
the black hole phase transition.  This result is useful in itself.
Following the dynamics of the black hole phase transition through a fully nonequilibrium 
formulation of semiclassical gravity is a very hard problem to which, at the moment, there is no direct method of attack.  
The study of simpler systems like the QBH 2LA with a similar entropic transition is likely to yield 
insight into the dynamics of the black hole phase transition.  In the next section we turn our attention to these
dynamical issues.

\section{Nonequilibrium dynamics}
\label{sec:dynamics}
Figure \ref{fig:bhF} appears qualitatively similar to the free energy of a 
system undergoing a (strongly) first-order phase transition.  The usual field theoretic treatment of the dynamics of
first-order phase transitions is based upon the homogeneous nucleation theory developed by Langer 
(see for example \cite{langer:1992}).  In
homogeneous nucleation, widely separated spherical bubbles of the stable phase nucleate in a background of the
unstable phase.  If the bubbles are larger than a critical radius, the volume
energy of the stable phase inside the bubble is less than the surface energy and the droplet will grow.  
The phase transition completes as droplets of the stable phase expand to fill the volume of the system.  In 
this picture,
the black hole phase transition occurs when an unstable black hole with mass $M_1$ nucleates 
(with probability $P \sim e^{-\beta F(M_1)}$) and grows to form the stable mass $M_2$ through the 
absorption of thermal radiation.  There are reasons to believe, however, that homogeneous nucleation is {\it not} the 
correct description, as we indicate below.

The free energy of the black hole system is calculated under the assumption of spherical symmetry, adequate 
only for a single 
black hole.  In equilibrium, a single black hole is preferred because it maximizes the entropy.   For example, 
in a state with
two black holes,
\begin{equation}
S_{m_a}+S_{m_b} \sim m_a^2+m_b^2 < S_{m_a+m_b} \sim (m_a+m_b)^2.
\end{equation}
However the dynamics of the phase transition may involve multiple black holes.  Black holes are strongly 
interacting gravitational systems, unscreened by thermal fluctuations.  It is possible that the dynamics of the
phase transition proceeds by the (exponentially more probable) nucleation of small black holes with mass $m \ll M_1$
which then merge to form larger holes. Phase transitions in which there are strong 
interactions between bubbles of the new phase lie beyond the scope of homogeneous nucleation.

To approach the dynamics of the black hole phase transition we study the dynamics of similar
physical systems.  The black hole phase transition is driven by the large amount of black hole
entropy $S \sim E^2$.  A thermodynamic relation where the entropy grows rapidly with energy is very unusual.
For example, a classical ideal gas has entropy $S\sim \ln{(E)}$.   However there are examples of
systems that exhibit similar entropic transitions. In particular we examine the Kosterlitz-Thouless (KT) transition 
in a global $O(2)$ model in $2+1$ dimensions \cite{kosterlitz:1973} and the Hagedorn transition 
\cite{hagedorn:1965} in string systems.  

\subsection{Kosterlitz-Thouless transition}
In two spatial dimensions, finite temperature fluctuations 
destroy long-range order in systems with broken continuous symmetries. Only at zero temperature does
the order parameter develop a non-zero thermal expectation value \cite{mermin:1966}.  However, at low
temperatures, systems with an O(2) symmetry do exhibit {\it algebraic} order: the
order parameter decays (in space) with a power-law, much like higher-dimensional systems near a critical point.
Since there is complete disorder at high temperatures (exponential decay of the order parameter)
there must be a transition from high temperature disorder to algebraic order.
This transition is unusual because it is driven by the fluctuations of vortex topological
defects.

To illustrate the KT transition we consider a global $O(2)$ scalar field model with Hamiltonian,
\begin{equation}
\label{eq-KTham}
H=\int d^2x \left ( \frac{1}{2}|\nabla \vec{\phi}|^2 + \frac{\lambda}{8} (\vec{\phi}^2-\eta^2)^2 \right ).
\end{equation}
This model admits vortex topological defects.  The energy of a vortex of single winding is
\begin{equation}
\label{eq-Ev}
E \approx \pi \eta^2 \left (\ln{\frac{R}{a}}+\lambda \eta^2 a^2 \right ),
\end{equation}
where $a \sim \frac{1} {\sqrt{\lambda} \eta}$ is the vortex size and $R$ is the size of the system.  If
$R \gg a$ the energy of the vortex is dominated by gradient energy, the first term in Eq.~(\ref{eq-Ev}).
If a vortex can be nucleated anywhere in the system then the entropy is
\begin{equation}
S=\ln{\left (\frac{R}{a}\right )^2},
\end{equation}
and the free energy of the system with a single vortex is
\begin{equation}
F = (\pi \eta^2 -2T)\ln{\left (\frac{R}{a} \right )}.
\end{equation}
Inspection of the free energy reveals that above the critical temperature 
\begin{equation}
T_c=\frac{\pi \eta^2}{2},
\end{equation}
it is thermodynamically favorable to nucleate vortices.  This is the KT transition.  It is the
proliferation of vortices that leads to the destruction of order above $T_c$. 
At low temperatures vortices exist as a low density of tightly bound
vortex-antivortex pairs with zero net topological charge.  As the temperature increases, both the density of
pairs and the average vortex-antivortex separation also increase.  
In addition, as the separation increases, thermal fluctuations are more effective in screening the 
interaction between defects in a vortex-antivortex pair. 
When the critical temperature is reached, screening is complete
and the vortex-antivortex pairs unbind leaving a gas of essentially free topological defects.

The vortex-driven KT transition provides a potentially useful framework for studying the the
black hole phase transition.  Analogous to the black hole transition, the nucleation of a vortex with energy $E$ produces a 
large amount of entropy $S \sim E$ which drives the phase transition.
The natural analogy of an antivortex in the black hole system is a {\it white} hole, a spacetime
region that expels all worldlines.  A white hole is the time-reverse of a black hole, just as 
(e.g. in superfluid helium) an antivortex is the time-reverse of a vortex.
In fact, a thermodynamic spacetime system obscures the differences between black holes and white holes:
an evaporating black hole looks like a white hole \cite{page:1981}.
The analogy of the KT transition suggests that the black hole phase transition might be more fully
understood by focusing upon the dynamical behavior of black holes and white hole in thermal
equilibrium.  Below the critical temperature of the black hole system we expect to find
an equal population of small (Planck scale) black holes and white holes continually fluctuating in
and out of existence.  The black holes result
from collapsed thermal radiation while the white holes result from evaporating black holes.
Perhaps the black hole phase transition is triggered by a slight overproduction of black holes which
then grow and coalesce to form the large black hole characteristic of the equilibrium state above
$T_c$. 

There are differences between the vortex and black hole system.  The vortex entropy results from
the possible locations of the vortex and not, as in the black hole, from internal states. In addition, in distinction to vortices, 
black holes and white holes do not carry opposite (gravitational) charge.  It is therefore likely that the 
mechanism driving the black hole phase transition is different in detail from
the unbinding of black hole-white hole pairs that we would expect if the KT analogy held exactly.  

The KT transition provides an excellent example of a dual theory.  The thermodynamics of 
the model with Hamiltonian (\ref{eq-KTham}) is in principle fully described by the O(2) field theory.  However, a full
characterization of the thermodynamic properties requires a nonperturbative analysis of the field theory (for
example through the use of lattice simulations).  In the dual vortex approach the system is
described as a Coulomb gas of charged vortices.  Because vortices are nonperturbative
field excitations, the dual vortex theory provides information complementary to the field order
parameter.  This information can be used to build a picture of the phase transition which (although completely
equivalent) is accessible only through a highly nonperturbative analysis of the field theory.  Unfortunately,
for the spacetime phase transition there is no rigorous technique for a nonperturbative analysis of the
underlying field theory of general relativity.    
It will take a more developed theory of quantum gravity than we presently have
in order to fully understand the dynamics of the black hole phase transition.  What our analysis {\it does} suggest is
that black holes and white holes may be the relevant degrees of freedom. Focusing upon them effectively
provides a new angle towards the dynamics of the black hole phase transition.  As an example, we provide below an explicit model
aimed at studying the dynamics of a black hole gas.

\subsubsection{Black hole gas}
To study the possibility that the dynamics of black hole phase transition proceeds through the
nucleation and merger of small black holes we propose a model in which black holes are treated
as a gas of gravitationally interacting particles. 
The number of particles is {\it not} fixed, but changes 
through the stochastic nucleation of small black holes, mergers and Hawking evaporation.  To incorporate the interaction of
black holes with the thermal environment the mass of each particle is not constant but changes in proportion to the free energy
of a single black hole,
\begin{equation}
\frac{dm(t)}{dt} \sim - \frac{\delta F}{\delta m}.
\end{equation}
>From $F(m)$ given by Eq.~(\ref{eq-bhf}) we obtain the phenomenological relation
\begin{equation}
\label{eq-massloss}
\frac{dm(t)}{dt}=-\frac{1}{\sqrt{1-\frac{2m}{r}}} +8\pi m T,
\end{equation}
where $T$ is the temperature and $r$ is the size of an effective boundary.
With a time-dependent mass given by Eq.~(\ref{eq-massloss}) black holes are stable
only above a critical temperature, as we expect from previous discussions.  The difference and advantage 
of this model is that above the critical temperature, a stable black hole can form by small mergers,
in addition to a single large fluctuation.   A conceptually similar approach but with
very different emphasis was considered in \cite{ellis:1999}. We hope to report on the analysis of this 
black hole gas model in the near future \cite{stephensB:2000}.

\subsection{Hagedorn transition}
A system of fundamental closed bosonic strings in thermodynamic equilibrium possesses an 
exponential density of states \cite{hagedorn:1965}
\begin{equation}
\label{eq-density}
\nu(E) \sim \exp{(\beta_H E)},
\end{equation}
where $\beta_H$ is the model and dimension dependent Hagedorn temperature.
Above the Hagedorn temperature the exponential density of states renders the
partition function ill-defined and a physical description of the system is unclear.  

As has been previously noted (see e.g \cite{bowick:1985}) the rapidly growing density
of states responsible for the Hagedorn transition is qualitatively similar to the large
black hole entropy driving the black hole phase transition.  In fact, recent advances in string
theory reveal a much stronger link between the Hagedorn transition and the formation of a black hole.

In string theory, the formation of a black hole in Anti-de Sitter (AdS) spacetime is linked through 
a hypothesized AdS/CFT correspondence to the Hagedorn transition in large-N 
SU(N) gauge theory (see \cite{aharony:2000} for a comprehensive review).
The thermodynamics of a semiclassical black hole system in a spacetime with negative 
cosmological constant is qualitatively similar to the black hole in a box discussed previously in Sec.~\ref{sec:thermo}.
In particular, a black hole phase transition, the Hawking-Page transition \cite{hawking:1983} occurs 
at a critical temperature 
\begin{equation}
T_c \sim \sqrt{|\Lambda |}
\end{equation}
where $\Lambda $ is the (negative) cosmological constant.
Exploiting the AdS/CFT correspondence, Witten argued \cite{witten:1998} that the enormous entropy released in 
the deconfinement transition of a supersymmetric gauge theory defined on the boundary of AdS 
spacetime has a dual bulk interpretation as a black hole forming through the Hawking-Page phase transition.
Insight into the interpretation of the Hagedorn transition is obtained by examining the relationship between 
string theory and Quantum Chromodynamics (QCD).  QCD is the (nonabelian) SU(3) gauge theory of strong interactions.
The fundamental degrees of freedom are three colored fermionic quarks interacting via bosonic gluons.  
QCD displays asymptotic freedom: at high energies the QCD coupling is weak and quarks are effectively free. 
However, at low energies QCD is strongly coupled and quarks exist only in tightly bound color singlet states, 
hadrons and mesons.  Between the two regimes, QCD is expected to undergo a deconfinement
phase transition at which hadrons and mesons melt into their constituent quarks.
The large-N \cite{thooft:1974} and strong coupling \cite{wilson:1974} limits of QCD may be described 
by a fundamental string
theory and the link between QCD and string theory provides a physical picture of the Hagedorn transition:
the proliferation of string states at the Hagedorn temperature corresponds to the 
emergence of quark color degrees of freedom at deconfinement \cite{pisarski:1982,olesen:1985,salomonson:1986}.
Through the AdS/CFT correspondence the deconfinement phase transition
corresponds to the Hawking-Page black hole phase transition.  These arguments present
the possibility that the dynamics of the Hagedorn transition is similar to the dynamics 
of black hole formation in the Hawking-Page transition.  

It is not clear off hand which  is more tractable: the dynamics of the Hagedorn transition in
fundamental string theory or the dynamics of the black hole phase transition.
However, the exponential density of states, Eq.~(\ref{eq-density}), also occurs in {\it classical}
string systems (see e.g \cite{sakellariadou:1996,smith:1987}).  
The advantage of a classical system is that the full nonperturbative dynamics is
accessible (albeit often only through numerical simulations).  
In a classical string system, the Hagedorn transition corresponds
to the abrupt formation of infinite string.  To date, only the equilibrium properties of the
Hagedorn transition in classical string systems have been explored.  We propose to study the nonequilibrium dynamics of this
transition \cite{stephensC:2000}. As a start, not only is it an interesting physical problem in its own right, it may also provide a tangible analogy for the dynamics of the black hole 
phase transition through the  correspondence mentioned above.  Much as the dynamics of a first or second-order phase
transition (nucleation or spinodal decomposition) is described independently of the particular system
under investigation, the dynamics of the Hagedorn transition may have similar universal
traits worthy of our perusal.

\section{Summary}
\label{sec:sum}
In these notes we have explored the basic issues and proposed some novel mechanisms in 
black hole phase transitions in semiclassical gravity.  We reviewed the semiclassical thermodynamics of the black hole system and
determined that the black hole phase transition is entropically driven: it
occurs because the large entropy of a black hole compensates for the energy cost of formation and
lowers the total free energy of the system.  This view was reinforced through the analysis of an 
atomic model of a quantum black hole in thermal equilibrium.  In this model, the phase
transition occurs as the abrupt excitation of a high energy state above the critical temperature.  Thus, the
study of atomic emission and absorption in this black hole atom may provide intuition about the black hole phase transition.
The detailed {\it dynamics} of the black hole 
phase transition remain an open problem.  We argued that homogeneous nucleation does not apply to phase transitions
in which there are strong interactions among bubbles of new phase, as is the case for the
black hole system.  We then appealed to the study of similar entropic transitions 
and observed that the black hole phase transition contains elements of both the Kosterlitz-Thouless and Hagedorn phase 
transitions.  Reasoning by analogy, we argued that the dynamics of the
black hole phase transition might be similar to the dynamical formation of long string
in the nonequilibrium quench of a classical string system.  
Work in these directions is in progress.

\bigskip
\noindent {\bf Acknowledgement}: This work is supported in part by NSF grant PHY98-00967.

\bibliography{blackhole}

\end{document}